\documentclass[final,5p,times,twocolumn]{elsarticle} 

\usepackage{lineno,hyperref}

\begin{document}

\begin{frontmatter}

\title{Progress in the development of a KITWPA for the DARTWARS project}

\author[unimib,infn-mi]{M. Borghesi\corref{mycorrespondingauthor}}
\cortext[mycorrespondingauthor]{Corresponding author. Email: matteo.borghesi@mib.infn.it}

\author[unisa,infn-sa]{C. Barone}
\author[unimib,infn-mi]{M. Borghesi}
\author[unimib,infn-mi]{S. Capelli}
\author[unisa,infn-sa]{G. Carapella}
\author[unisalento,infn-le]{A. P. Caricato}
\author[ino-cnr,unitn]{I. Carusotto}
\author[fbk,infn-tifpa]{A. Cian}
\author[lnf]{D. Di Gioacchino}
\author[inrim,infn-tifpa]{E. Enrico}
\author[fbk,infn-tifpa,ifn-cnr]{P. Falferi}
\author[inrim,polito]{L. Fasolo}
\author[unimib,infn-mi]{M. Faverzani}
\author[infn-mi]{E. Ferri}
\author[infn-sa,unisannio-sci]{G. Filatrella}
\author[lnf]{C. Gatti}
\author[unimib,infn-mi]{A. Giachero}
\author[fbk,infn-tifpa]{D. Giubertoni}
\author[unisa,infn-sa]{V. Granata}
\author[inrim,polito]{A. Greco\fnref{myfootnote}}
\author[unisa,infn-sa]{C. Guarcello}
\author[unimib,infn-mi]{D. Labranca}
\author[unisalento,infn-le]{A. Leo}
\author[lnf]{C. Ligi}
\author[lnf]{G. Maccarrone}
\author[fbk,infn-tifpa]{F. Mantegazzini}
\author[fbk,infn-tifpa]{B. Margesin}
\author[unisalento,infn-le]{G. Maruccio}
\author[infn-sa]{C. Mauro}
\author[unitn,infn-tifpa]{R. Mezzena}
\author[unisalento,infn-le]{A. G. Monteduro}
\author[unimib,infn-mi]{A. Nucciotti}
\author[inrim,infn-tifpa]{L. Oberto}
\author[unimib,infn-mi]{L. Origo}
\author[unisa,infn-sa]{S. Pagano}
\author[infn-sa,unisannio-eng]{V. Pierro}
\author[lnf]{L. Piersanti}
\author[inrim,infn-to]{M. Rajteri}
\author[lnf]{A. Rettaroli}
\author[unisalento,infn-le]{S. Rizzato}
\author[fbk,infn-tifpa,ifn-cnr]{A. Vinante}
\author[unimib,infn-mi]{M. Zannoni}

\address[lnf]{INFN - Laboratori Nazionali di Frascati, Via Enrico Fermi, I-00044, Frascati, Italy}
\address[infn-le]{INFN – Sezione di Lecce, Via per Arnesano, I-73100 Lecce, Italy}
\address[unisalento]{University of Salento, Department of Physics, Via per Arnesano, I-73100 Lecce, Italy}
\address[unimib]{University of Milano Bicocca, Department of Physics, Piazza della Scienza , I-20126 Milano, Italy}
\address[infn-mi]{INFN - Milano Bicocca, Piazza della Scienza, I-20126 Milano, Italy}
\address[unisa]{University of Salerno, Department of Physics, Via Giovanni Paolo II, I-84084 Fisciano, Salerno, Italy}
\address[infn-sa]{INFN - Napoli, Salerno group, Via Giovanni Paolo II, I-84084 Fisciano, Salerno, Italy}
\address[unisannio-sci]{University of Sannio, Department of Science and Technology, via Francesco de Sanctis, I-82100, Benevento, Italy}
\address[unisannio-eng]{University of Sannio, Department of Engineering, Corso Garibaldi, I-82100 Benevento, Italy}
\address[ino-cnr]{INO-CNR BEC Center,  Via Sommarive, I-38123 Povo, Italy}
\address[unitn]{University of Trento, Department of Physics, Via Sommarive, I-38123, Povo, Trento, Italy}
\address[fbk]{Fondazione Bruno Kessler, Via Sommarive, I-38123, Povo, Trento, Italy}
\address[infn-tifpa]{INFN - TIFPA, Via Sommarive, I-38123, Povo, Trento, Italy}
\address[ifn-cnr]{IFN-CNR , Via Sommarive, I-38123 Povo, Trento, Italy}
\address[inrim]{INRiM - Istituto Nazionale di Ricerca Metrologica, Strada delle Cacce, I-10135 Turin, Italy}
\address[polito]{Politecnico di Torino, Corso Duca degli Abruzzi, I-10129 Turin, Italy}
\address[infn-to]{INFN - Torino, Via Pietro Giuria, I-10125 Turin, Italy}





\begin{abstract}
DARTWARS (Detector Array Readout with Traveling Wave AmpliﬁeRS) is a three years project that aims to develop high-performing innovative Traveling Wave Parametric Ampliﬁers (TWPAs) for low temperature detectors and qubit readout (C-band). The practical development follows two diﬀerent promising approaches, one based on the Josephson junctions (TWJPA) and the other one based on the kinetic inductance of a high-resistivity superconductor (KITWPA). This paper presents the advancements made by the DARTWARS collaboration to produce a first working prototype of a KITWPA.
\end{abstract}


\end{frontmatter}


\section{Introduction}
Nowadays, many applications rely on the faithful detection of low power microwave signals at cryogenic temperatures. This is especially true in the field of the superconducting quantum computation, where quantum-limited noise microwave amplification is paramount to infer the qubit state with high fidelity.

For these applications, the goals are to maximize the signal to noise ratio of extremely feeble microwave signals while allowing a broad readout bandwidth. The latter is extremely relevant in all the applications where the devices are required to be multiplexed over large bandwidths.
To solve this problem, parametric amplification with superconducting circuits, a well known technique used for low noise amplifiers, will be exploited and developed to its technical limits.

DARTWARS \cite{giachero2021detector} (Detector Array Readout with Traveling Wave AmplifieRS) aims to develop two types of traveling wave parametric amplifiers: TWJPA and KITWPA. The technical goal is to achieve a gain value around 20 dB, such that the added noise of a commercial semiconductors low temperature amplifiers (HEMT) could be neglected, with a high saturation power (around -50 dBm), and a quantum limited or nearly quantum limited noise ($T_N < 600$ mK) in C-band. These features will lead to the readout of large arrays of detectors or qubits with virtually no noise degradation.

\section{Parametric amplification and KITWPA} 

A traveling wave parametric amplifier is the physical implementation of a parametric oscillator. By carefully modulating the properties of the oscillating system, the oscillator will absorb energy from the surroundings, resulting in an exponential growth of the oscillation amplitudes.
A parametric amplification of a current signal $I$ can  thus be achieved by making the signal travel through a transmission line (TL) made of a non-linear inductors $L(I)$, resulting in a wave equation which is equal to the one of the parametric oscillator.
In particular, KITWPA exploits the nonlinear kinetic inductance of a superconducting coplanar waveguide (CPW) to generate parametric interaction between a strong RF input pump at a frequency $\omega_p$ and signal at a frequency $\omega_s$ ($\omega_s < \omega_p$) ,  resulting in a signal amplification and generation of idler(s) product(s) $\omega_i$ \cite{erickson2017theory}.

The inductance of the superconducting line comprises a geometric inductance $L_g$ and a kinetic inductance $L_{k}$. The former depends on the geometry of the circuit, while the latter depends on the supercunducting material and it is related to the inertia of the Cooper pairs. The kinetic inductance makes the overall inductance of the line non linear, and along the direction $x$ of the CPW can be modeled as 
\begin{equation}\label{eq-I_star}
L(x,t) = L_0 (x) \Bigg[ 1 + \Big( \frac{I(x,t)}{I^*} \Big)^2    \Bigg]
\end{equation}
where $L_0(x)$ is the linear kinetic inductance and $I^*$ is a constant scaling factor.

The fact that a superconducting line has zero $dc$ resistance at $T<T_c$ suggests the motivation for the use of superconducting parametric amplifiers: a very low power dissipation and an added noise that approaches the minimum set by quantum mechanics.

One of the goal of DARTWARS is to develop a KITWPA amplifiers that operate in a three-wave mixing (3WM) fashion when biased with a dc current $I_d$. In this scheme, the pump exchange its energy with a signal and idler products when the following phase-matching conditions are satisfied

\begin{equation}\label{eq-phase_match}
\omega_p = \omega_s + \omega_i  \\ ;\\  k_p-k_s-k_i = \frac{\chi I_{p0}^2}{8}(k_p – 2k_s -2k_i)
\end{equation} 
where ${k_p,k_s,k_i}$ are the wavenumbers of the corresponding signals and $\chi = 1/(I^{*2}+I_d^2)$ and $I_{p0}$ is the pump amplitude at the amplifier input.


\section{Preliminary results from the KITWPA fabrication}

To achieve the phase matching condition of eq (\ref{eq-phase_match}), DARTWARS will exploit a lumped-element transmission line for the KITWPA design. In this configuration, the characteristic impedance of the TL is periodically loaded to implement  both a wide stopband at $3\omega_p$ and a narrow stopband near $\omega_p$. Placing the pump tone below this narrow stopband allows to easily match the conditions of eq (\ref{eq-phase_match}) while suppressing higher pump harmonics, which would result in a gain and noise degradation. In addition, the lumped elements TL allows to achieve a high gain with a shorter TL (tens of cm compared to few meters of a periodic loading TL).

\begin{figure}[h!]
	\centering
	\includegraphics[width=1\linewidth]{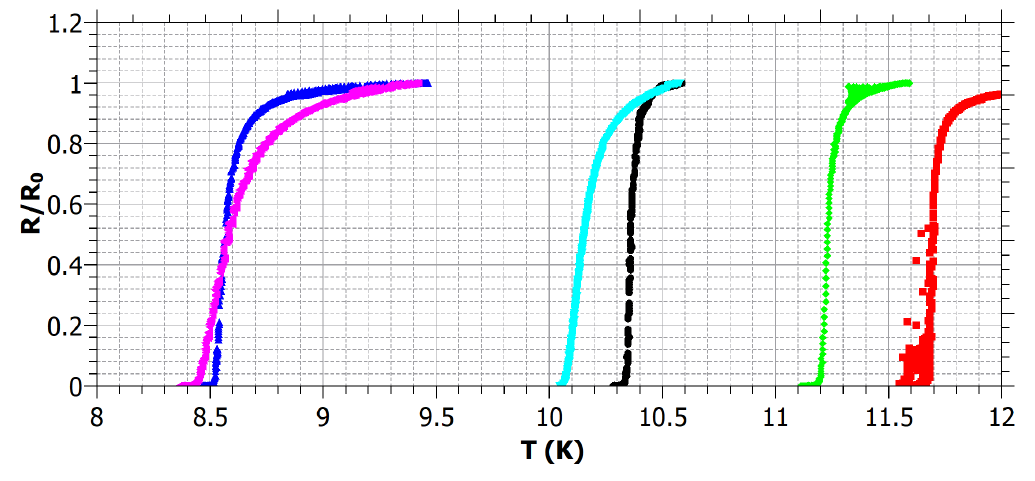}
	\caption{Normalized transition curves of the NbTiN films fabricated at FBK. The different colors correspond to different film depositions.}
	\label{fig:Tc}
\end{figure}

The material of choice to fabricate the KITWPAs is NbTiN and the films are being produced at the FBK facility. The critical temperature of the film produced can be seen in Figure \ref{fig:Tc}. 

Even if the measured critical temperatures suggest that new sputter targets will be required to achieve the best literature values ($\sim 15$ K), the results of the sputtering fabrication processes are encouraging, indicating that a wide range of $L_k$ can be easily achieved with the current setup.

\begin{figure}[h!]
	\centering
	\includegraphics[width=0.9\linewidth]{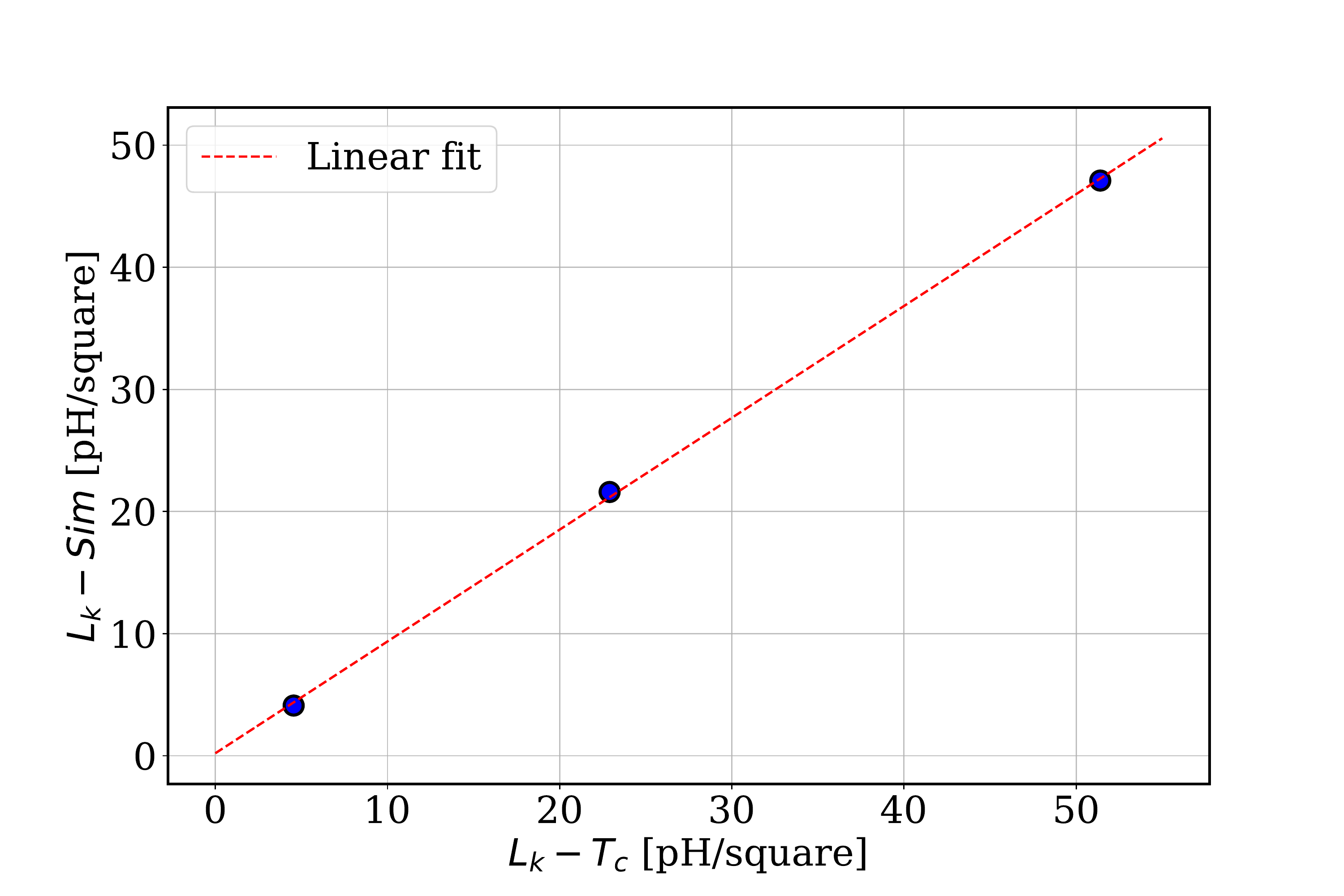}
	\caption{Comparison between the two different methods adopted to measure the kinetic inductance of the supeconducting films.}
	\label{fig:Lk}
\end{figure}

To characterize the produced films, the NbTiN has been patterned into lumped element resonators \cite{day2003broadband}. The kinetic inductance was assessed on three different wafers with two techniques. The first technique, labeled as $L_k$-$Sim$, estimates $L_k$ as the value required to match the measured resonant frequency of the KID with the one extracted from simulation, which takes into account only $L_g$. The second approach, indicated as $L_k$-$T_c$, computes $L_k$ from 
\begin{equation}
L_k = \frac{\hbar}{1.76 \pi k_b T_c}R_n
\end{equation}  
The two techniques are comparable, as shown in Figure \ref{fig:Lk}. 

\begin{figure}[h!]
	\centering
	\includegraphics[width=0.9\linewidth]{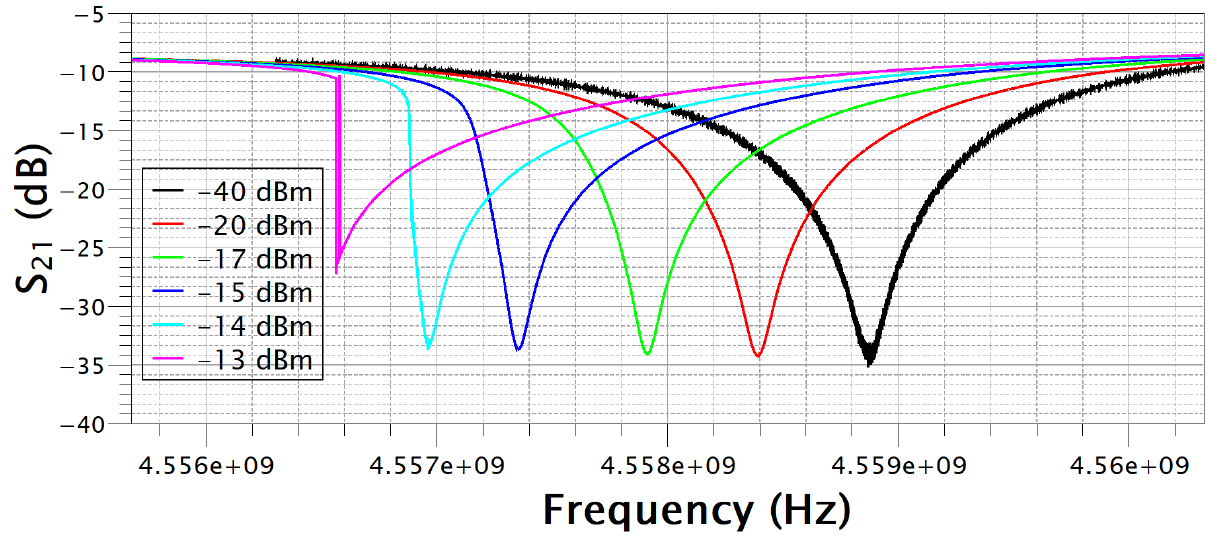}
	\caption{Example of different resonant profiles due to the different powers of the RF signal.}
	\label{fig:res_power}
\end{figure}

 Finally, the maximum achievable non-linearity of the film $\frac{I_c}{I^*}$ was assessed from eq (\ref{eq-I_star}). The resonance frequency $f_0$ was measured by changing the power $P$ of the RF probe tones (Fig \ref{fig:res_power}), while the conversion between $P$ and $I^2$ was assessed through Sonnet and Qucs simulations. The following relation  was used to evaluate $L_k$: $(2\pi f_0)^{-2} = (L_k + L_g)C$.
 
\begin{figure}[h!]
	\centering
	\includegraphics[width=0.9\linewidth]{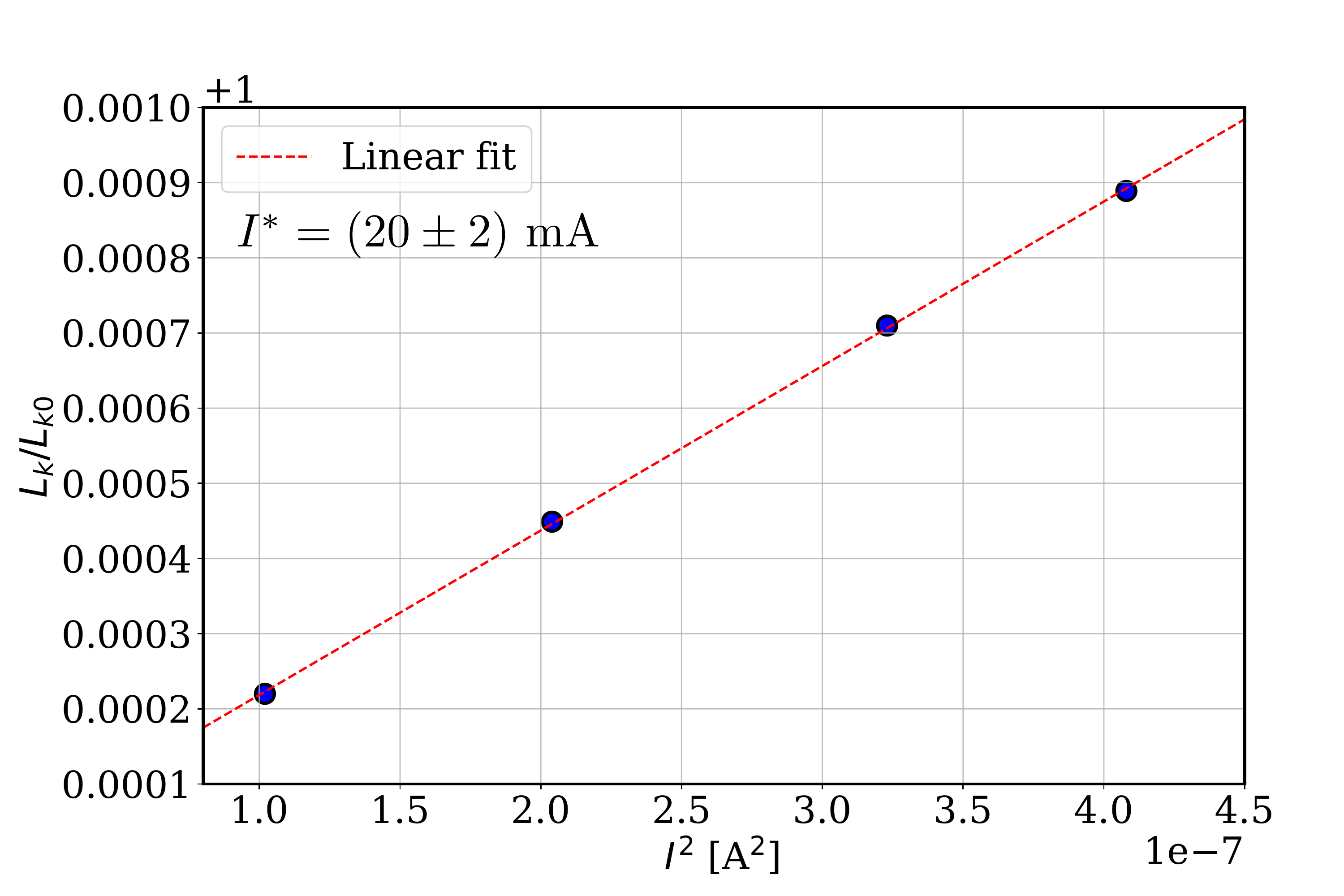}
	\caption{$I^*$ measurement of the resonator with the highest kinetic inductance}
	\label{fig:I_star}
\end{figure}

Figure \ref{fig:I_star} shows the results achieved for one of the produced resonators. Overall, the maximum achieved non-linearity from the different wafers produced was roughly 0.25, close to the literature value of 0.34 \cite{malnou2021three}. 
Upon these promising results, the first KI-TWPA prototypes are currently beeing produced.

\section*{Acknowledgments}
This work is supported by the Italian Institute of Nuclear Physics (INFN), within the Technological and Interdisciplinary research commission (CSN5), by the European Union's H2020-MSCA Grant Agreement No. 101027746, and by the Joint Research Project PARAWAVE of the European Metrology Programme for Innovation and Research (EMPIR). PARAWAVE received funding from the EMPIR programme co-financed by the Participating States and from the European Union's Horizon 2020 research and innovation programme.




\bibliography{mybibfile}

\end{document}